\documentclass[aps, preprint,superscriptaddress, prl,showpacs,showkeys,floatfix]{revtex4-1}

\usepackage{amsmath}
\usepackage{amssymb}
\usepackage{bm}
\usepackage{graphicx}
\usepackage{xcolor}

\newcommand{\bmath}{\begin{mathletters}}
\newcommand{\emath}{\end{mathletters}}
\newcommand{\be}{\begin{eqnarray}}
\newcommand{\ee}{\end{eqnarray}}
\newcommand{\ba}{\begin{array}}
\newcommand{\ea}{\end{array}}

\providecommand\ket[1]{\left\lvert#1\right\rangle}

\providecommand\ketbra[2]{\left\lvert{#1}\middle\rangle\!\middle\langle{#2}\right\rvert}
\providecommand\braket[2]{\left\langle{#1}\middle\vert{#2}\right\rangle}

\providecommand\si[1]{\ensuremath{\,\mathrm{#1}}}
\providecommand\SI[2]{\ensuremath{#1\,\mathrm{#2}}}

\newcommand{\rmd} {{\rm d}}

\begin{document}

\title{Measuring the Berry Phase in a Superconducting Phase Qubit by a Shortcut to Adiabaticity }
\author{Zhenxing Zhang}
\author{Tenghui Wang}
\author{Liang Xiang}
\author{Jiadong Yao}
\author{Jianlan Wu}
\affiliation{Department of Physics, Zhejiang University, Hangzhou 310027, China}
\author{Yi Yin}
\email{yiyin@zju.edu.cn}
\affiliation{Department of Physics, Zhejiang University, Hangzhou 310027, China}
\affiliation{Collaborative Innovation Center of Advanced Microstructures, Nanjing 210093, China}


\clearpage

\begin{abstract}
With a counter-diabatic  field  supplemented to the reference control field, the `shortcut to adiabaticiy' (STA) protocol is implemented in a superconducting phase qubit.
The Berry phase measured in a short time scale
is in good agreement with the theoretical result acquired from an adiabatic loop.
The trajectory of a qubit vector is extracted, verifying
the Berry phase alternatively by the integrated solid angle.
The classical noise is introduced to the amplitude or phase of the total control field.
In the statistics of the Berry phase, the mean with either noise
is almost equal to that without noise,
while the variation with the amplitude noise can be described by an analytical expression.
\end{abstract}

\maketitle

In quantum mechanics, a geometric phase is acquired as a quantum state propagates
on a curved surface and this interesting phenomenon is observed in many
quantum systems ranging from microscopic particles to condensed matter materials~\cite{BerryPhase,Chiao,Bitter,WilczekBook,BerrySolidState}.
The precise control and measurement of the geometric phase lead to geometric phase gates
in quantum computation~\cite{SjoqvistPhysics,NMRgate,Iongate,nonAbelianGate,SpinGate}.
A typical example to understand the geometric phase is a spin-$1/2$ particle subject to a slowly-varying
magnetic field which undergoes a closed path in the parameter space. In response
to the changing field, the spin remains in an instantaneous eigenstate and
follows  a cyclic path in the Bloch sphere.
The geometric phase, termed the Berry phase~\cite{BerryPhase},
is acquired by  the spin state when the spin vector returns to its initial position.
The Berry phase in   such an adiabatic process
is independent of the speed of the field's evolution
if the adiabatic condition is satisfied.

In practice, a quantum manipulation is often performed in a short time scale to avoid dissipation induced errors,
incompatible with the presumption of the adiabatic process.
The advance of large-scale quantum devices requires fast operations to
improve efficiency of information processing~\cite{XMon2}. Various procedures have
thus been proposed for the realization of  a fast `adiabatic' process~\cite{DemirplakRice,Berry09,ChenXi,Masuda,Shortcuts,CompositPulse,fastAdiabaticgate}.
One general strategy is to apply a `shortcut to adiabaticity' (STA)
where an additional Hamiltonian is employed to cancel the non-adiabatic contribution in a fast evolution~\cite{DemirplakRice,Berry09,ChenXi,Masuda,Shortcuts}.
For a given reference Hamiltonian $H_0(t)$, the counter-diabatic Hamiltonian $H_{\mathrm{cd}}(t)$ is formally written as~\cite{Berry09,SI}
\be
H_{\mathrm{cd}}(t) = i\hbar\sum_n \big[\ketbra{\partial_t n(t)}{n(t)}-\braket{n(t)}{\partial_t n(t)}\ketbra{n(t)}{n(t)}\big],
\label{eq:1}
\ee
where $|n(t)\rangle$ is an instantaneous eigenstate of $H_0(t)$.
For each eigenstate $|n(t)\rangle$, $H_{\mathrm{cd}}(t)$
suppresses its non-adiabatic transition to other eigenstates.
The quantum system driven
by the total Hamiltonian $H_\mathrm{tot}(t)=H_0(t)+H_{\mathrm{cd}}(t)$ can evolve fast but
remain in the state $\ket{n(t)}$.
Regardless of the evolution of $H_\mathrm{tot}(t)$, the system state
evolves along a closed path with respect to $H_0(t)$ and acquires
the associated Berry phase~\cite{AAphase,ZhuSL2002}.

The STA protocol has been implemented in atomic, molecular and optical systems
for the state preparation, population transfer and optimal control~\cite{Shortcuts,Bason,Suter}.
Compared with these microscopic systems, a superconducting circuit is
fabricated on chips with lithographic scalability.
The superconducting qubit is realized based on nonlinear quantized energy levels of the circuit.
Sophisticated microwave techniques allow a reliable generation of the counter-diabatic Hamiltonian
in superconducting qubits. In this paper, we focus on
the realization of the STA protocol in a superconducting phase
qubit~\cite{MartinisReview,phaseQubit}. We achieve
the Berry phase measurement with the STA protocol and
study the influence of external field fluctuations,
which extends previous studies of the Berry phase in a Cooper pair pump~\cite{MottononPhase}
and in an adiabatically-steered superconducting qubit~\cite{LeekPhase,NoisePhase} .

The circuit of a superconducting phase qubit is a nonlinear resonator
comprised of a Josephson junction, a loop inductance and a
capacitor~\cite{SI,MartinisReview,phaseQubit}. A flux current biases this resonator
in an anharmonic cubic potential, and the lowest two energy levels of the nonlinear resonator
are used as the ground ($\ket{0}$) and excited  ($\ket{1}$) states of a qubit. Through the same bias line,
a microwave drive signal
is coupled to the qubit, providing a fast and reliable control of the qubit state.
The phase qubit is mounted in a sample box and cooled in a dilution refrigerator
whose base temperature is $\sim 10$ mK.
In our experiment, the resonance frequency of the phase qubit is set at $\omega_{10}/2\pi=5.7\si{GHz}$,
and the qubit dissipation is characterized by  a relaxation time, $T_1=\SI{270}{ns}$, and a spin-echo
decoherence time, $T_2^{\mathrm{echo}}=\SI{450}{ns}$.
For our phase qubit, it is difficult to perform an
adiabatic operation~\cite{LeekPhase,NoisePhase},
but feasible to implement a fast
STA protocol. With the qubit modelled as a spin-$1/2$ particle, we also treat
 the microwave signal as an effective external magnetic field.
In the rotating frame of the external field, the Hamiltonian is expressed as
 $H(t)= \hbar\bm{B}(t)\cdot\bm{\sigma}/2$ after the rotating wave approximation~\cite{SI}.
Here $\bm{\sigma}= (\sigma_x, \sigma_y, \sigma_z)$ is
the vector of Pauli operators and $\bm{B}(t)$ is the effective magnetic field
expressed in the unit of angular frequency.
Throughout this paper, our experiment will be described and discussed in the rotating frame.

\begin{figure}
\centering
\includegraphics[width=0.65\columnwidth]{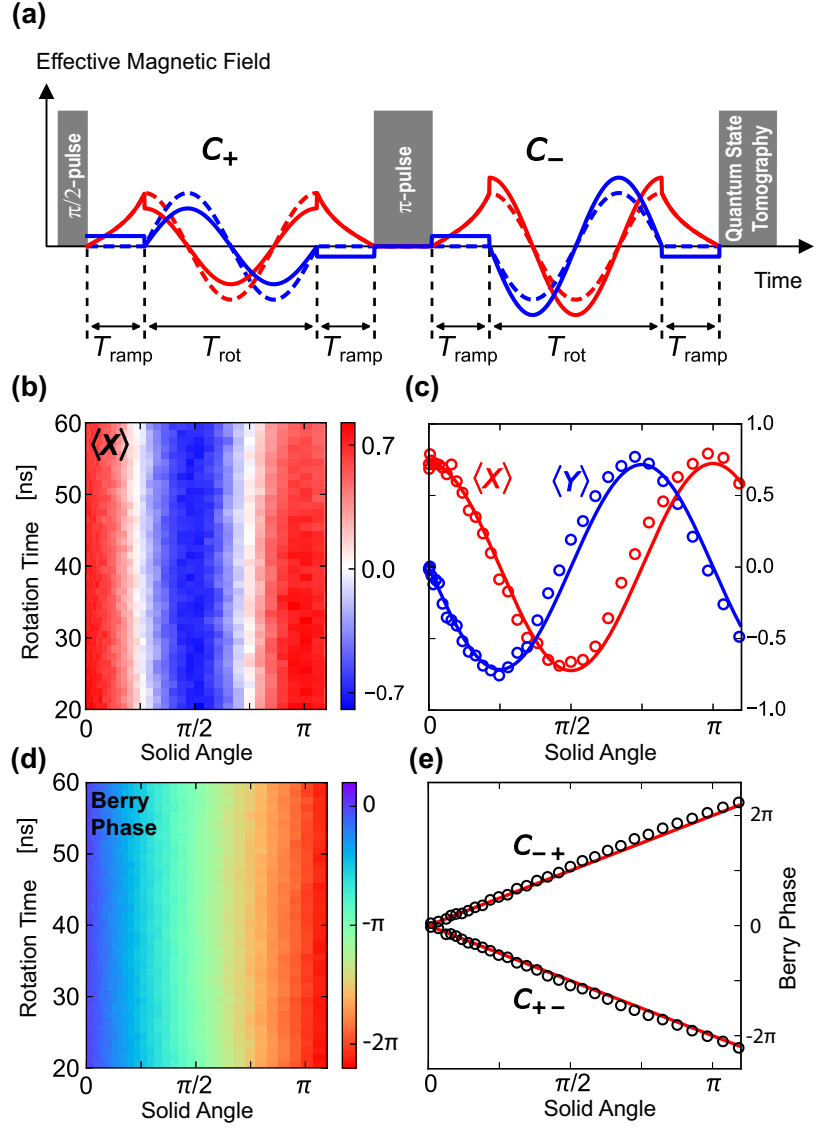}
\caption{ The STA experiment of measuring the Berry phase with a $\mathcal C_{+-}$ spin-echo procedure.
(a) The schematic diagram of the effective magnetic field in the $x$-$y$ plane.
The dashed and solid lines are the reference and total control fields,
while the red and blue colors refer to their $x$- and $y$-components.
(b) The measurement of the $x$-projection of the final qubit state versus
the solid angle $\mathcal S$ and the rotation time $T_\mathrm{rot}$.
(c) With $T_\mathrm{rot}=30$ ns, the $x$-, $y$-projections of the final qubit state (red and blue circles) are compared with the theoretical prediction (red and blue lines).
(d) The measurement of the Berry phase versus $\mathcal S$ and $T_\mathrm{rot}$.
(e) With $T_\mathrm{rot}=30$ ns, the extracted Berry phase (circles) is compared with the theoretical prediction (solid line). The comparison between the experimental measurement
and the theoretical prediction in the $\mathcal C_{-+}$ procedure
is also presented.
}
\label{fig:fig1}
\end{figure}

In an adiabatic process, the Berry phase can be measured with a spin-echo scheme~\cite{LeekPhase,NoisePhase,FilippPhase}.
In our STA experiment, the reference Hamiltonian, $H_0(t)=\hbar\bm B_0(t)\cdot\bm\sigma/2$,
is used to construct the spin-echo trajectory in a short time scale.
As shown in Fig.~\ref{fig:fig1}(a), the reference magnetic field $\bm B_0(t)$ evolves as follows.
A $\pi/2$-pulse is applied to the ground-state qubit,
preparing an initial superposition state at $(\ket{0}+\ket{1})/\sqrt{2}$.
After the initialization, the first sequence of the field,
$\bm{B}_{\mathrm{ramp},0}(t_1)=(\Delta_0\tan\theta(t_1), 0, \Delta_0)$ with $0<t_1<T_\mathrm{ramp}$,
ramps up its $x$-component by $\theta(t_1)=\theta_0 t_1/T_\mathrm{ramp}$.
The ramping time is fixed at $T_\mathrm{ramp}=10$ ns,
while  $\Delta_0 = \omega_\mathrm{d} - \omega_{10}$   is the detuning between the microwave drive frequency $\omega_\rmd$
and the qubit resonance frequency $\omega_{10}$. We set $\Delta_0/2\pi=7$ MHz
to reduce the influence of higher excited states.
The second sequence builds a rotating field,
$\bm B_{\mathrm{rot}, 0}(t_2)=(\Omega_0 \cos \phi(t_2), \Omega_0 \sin \phi(t_2), \Delta_0)$ with $0<t_2<T_\mathrm{rot}$.
The drive amplitude is $\Omega_0=\Delta_0\tan\theta_0$,
while the rotation is operated under a constant speed $\omega_0=2\pi/T_\mathrm{rot}$
along either the counterclockwise ($\mathcal{C}_+$) or clockwise ($\mathcal{C}_-$) direction.
The time-dependent phase of the effective field is given by $\phi(t_2)=\pm \omega_0 t_2$.
In our experiment,  the pre-designed polar angle $\theta_0$ and the rotation time $T_\mathrm{rot}$ are varied as control parameters.
To further reduce the influence of higher excited  states, a reversed ramping-down
field, $\bm{B}_{\mathrm{ramp},0}(T_\mathrm{ramp}-t_3)$ with $0<t_3<T_\mathrm{ramp}$,
is subsequently used to finish the first (dephasing) part of the spin-echo scheme~\cite{3level}.
A refocusing $\pi$-pulse is then applied to invert the qubit states.
During the second (rephasing) part of our spin-echo scheme, the three sequences in
the first part are reversed, as shown in Fig.~\ref{fig:fig1}(a).
Based on the cyclic rotations in the dephasing and rephasing parts,
the symbols of $\mathcal{C}_{+-}$ and $\mathcal{C}_{-+}$ represent two different types of spin-echo procedures.

To fulfill the STA protocol, the counter-diabatic Hamiltonian,
$H_{\mathrm{cd}}(t)=\hbar \bm B_{\mathrm{cd}}(t)\cdot\bm\sigma/2$,
is calculated by Eq.~(\ref{eq:1}). For each ramping or rotating step,
the counter-diabatic magnetic field is given by $\bm B_{\mathrm{cd}}(t)=\bm B_0(t)\times\dot{\bm B}_0(t)/|\bm B_0(t)|^2$,
which is perpendicular to the reference magnetic field~\cite{SI}.
In particular, the counter-diabatic ramping field is written as
$\bm B_{\mathrm{ramp}, \mathrm{cd}}(t)=(0, \pm\theta_0 /T_\mathrm{ramp}, 0)$,
where the $\pm$ signs correspond to the ramping-up and ramping-down steps, respectively.
The counter-diabatic rotating field is written as
$\bm{B}_\mathrm{rot,cd}(t) = \left(\Omega_\mathrm{cd}\cos\phi(t), \Omega_\mathrm{cd}\sin\phi(t),
\Delta_\mathrm{cd}\right)$, where the signs of
$\Omega_{\mathrm{cd}}=\mp\omega_0\sin\theta_0\cos\theta_0$
and $\Delta_\mathrm{cd}=\pm\omega_0\sin^2\theta_0$ refer to the
$\mathcal{C}_+$ and $\mathcal C_-$ rotations, respectively.
The total magnetic field is obtained as $\bm B_\mathrm{tot}(t)=\bm B_0(t)+\bm B_{\mathrm{cd}}(t)$,
an example of which is shown in Fig.~\ref{fig:fig1}(a).

In the STA protocol, the qubit is designed to follow the path of $\bm B_0(t)$ when driven by
the total Hamiltonian $H_\mathrm{tot}(t)=\hbar\bm B_\mathrm{tot}(t)\cdot \bm\sigma/2$.
Initially, the $|0\rangle$ and $|1\rangle$ states
are the instantaneous spin-up ($\ket{s_\uparrow(t)}$) and spin-down ($\ket{s_\downarrow(t)}$) eigenstates  of the reference Hamiltonian, respectively.
In an ideal adiabatic scenario, the Bloch vector of the $\ket{s_\uparrow(t)}$ state always points to the same direction of $\bm B_0(t)$,
while the opposite occurs for the $\ket{s_\downarrow(t)}$ state. The ramping sequences only produce a dynamic phase,
while both the dynamic and Berry phases are acquired during circular rotations. For the $\ket{s_\uparrow(t)}$ state, the Berry phase accumulated in
one cycle is given by $\gamma_{\uparrow}=\mp\mathcal S/2$, where $\mathcal S=2\pi(1-\cos\theta_0)$ is
the solid angle of the cone subtended by the cyclic path at the origin and the $\mp$ signs refer to the $\mathcal{C}_+$ or
$\mathcal{C}_-$ paths, respectively. Since the $\ket{s_\downarrow(t)}$ state follows the opposite path of
the $\ket{s_\uparrow(t)}$ state, its accumulated Berry phase is opposite, i.e., $\gamma_{\downarrow}=\pm\mathcal S/2$.
After the refocusing $\pi$ pulse, the phases associated with the $\ket{s_\uparrow(t)}$
and $\ket{s_\downarrow(t)}$ states are swapped. For each instantaneous eigenstate, the dynamic phase is cancelled in the rephasing part,
while the Berry phase is doubled due to a reversed rotating direction.
At the echo time, the $\ket{s_\uparrow(t)}$  and $\ket{s_\downarrow(t)}$ states return to the two poles of the Bloch sphere,
and the final qubit state is written as $(e^{-i\gamma/2}\ket{0}+e^{i\gamma/2}|1\rangle)/\sqrt{2}$ with a global
phase excluded~\cite{SI}. The total relative Berry phase after two circular rotations
is theoretically given by $\gamma= \mp 2\mathcal S$ for the $\mathcal{C}_{+-}$ or $\mathcal{C}_{-+}$ procedures, respectively.

In our experiment, the total field $\bm B_\mathrm{tot}(t)$ is realized by the microwave control signal,
and the qubit is driven in the spin-echo scheme as above.
To measure the Berry phase $\gamma$, we apply the quantum state tomography (QST) to extract the density matrix $\rho$ of the final state~\cite{phaseQubit}.
The $x$- and $y$-projections of the qubit vector are given by
$\langle x\rangle = \mathrm{Tr}\{\sigma_x \rho\}$ and
$\langle y\rangle = \mathrm{Tr}\{\sigma_y \rho\}$.
The two-dimensional (2D) diagram in Fig.~\ref{fig:fig1}(b) presents the experimental measurement of
$\langle x\rangle$ with the change of the rotation time $T_{\mathrm{rot}}$ and the pre-designed solid angle $\mathcal S$ (through the change of $\theta_0$)
in the $\mathcal C_{+-}$ procedure.
The rotation time, $20~\mathrm{ns}\le T_{\mathrm{rot}}\le 60~\mathrm{ns}$,
in our experiment is much shorter than the necessary time ($T_{\mathrm{rot}}\gtrsim 1~ \mu$s) of an adiabatic operation~\cite{LeekPhase} .
We observe that $\langle x\rangle$ oscillates with $\mathcal S$
and varies slowly with $T_\mathrm{rot}$;
the same conclusion is applied to $\langle y\rangle$ (not shown).
From the results of $T_\mathrm{rot}=30$ ns in Fig.~\ref{fig:fig1}(c), these two components
are consistent with the theoretical prediction of $\langle x\rangle =r\cos\gamma$ and $\langle y\rangle =r \sin\gamma$
with $\gamma=-2\mathcal S$.
Here $r=0.72$ is an adjusted fitting parameter
due to dissipation.
On the other hand, these two oscillations are nearly undamped with $\mathcal S$, indicating a weak geometric
dephasing~\cite{LeekPhase}. For each final density matrix, the Berry phase $\gamma$ is  estimated using $\arctan[\langle y\rangle/\langle x\rangle]$.
In the $\mathcal C_{+-}$ procedure, the Berry phase with $\theta_0<\pi/3$ (or $\mathcal S<\pi$)
is assigned in the range of $(-2\pi, 0)$ after considering the signs of $\langle x\rangle$ and $\langle y\rangle$.
With $\pi/3<\theta_0<\pi/2$ (or $\pi<\mathcal S<2\pi$), an extra $-2\pi$ is included to assign $\gamma$ in the range of $(-4\pi, -2\pi)$.
An opposite range is specified for $\gamma$ in the $\mathcal C_{-+}$ procedure.
Figure~\ref{fig:fig1}(d) presents the measurement of $\gamma$ for the $\mathcal C_{+-}$ procedure.
From the result of $T_\mathrm{rot}=30$ ns in Fig.~\ref{fig:fig1}(e),
the linear relation, $\gamma=-k\mathcal S$ ($k=2.04\pm0.02$),
is extracted, in good agreement with the theoretical prediction of $\gamma=-2\mathcal S$.
The same relation with an opposite sign, $\gamma=k^\prime \mathcal S$ ($k^\prime=2.06\pm0.02$),
is extracted for the $\mathcal C_{-+}$ procedure. The two slopes, $k$ and $k^\prime$, are almost unchanged
with the increase of the rotation time $T_\mathrm{rot}$.
Our experiment thus demonstrates that the Berry phase of a cyclic adiabatic path can be
successfully measured in the superconducting phase qubit following the fast STA protocol.

\begin{figure}
\centering
\includegraphics[width=0.65\columnwidth]{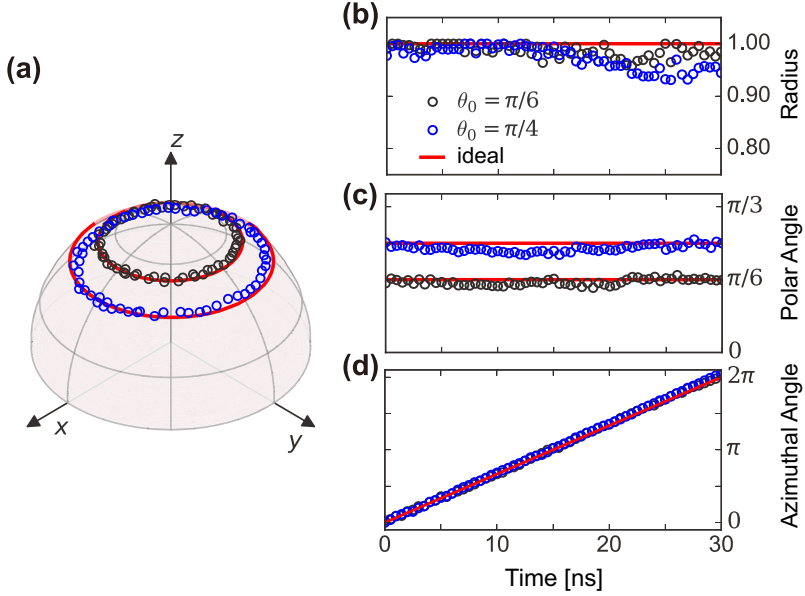}
\caption{In the STA procedure, the evolution of the $\ket{s_\uparrow(t)}$ state subject to a $\mathcal C_+$-rotating field with
$T_\mathrm{rot}=30$ ns.  (a) The trajectory of the qubit vector in the Bloch sphere.
In the spherical coordinate system of the qubit vector, the time evolutions of radius, polar
and azimuthal angles are plotted in (b), (c) and (d), respectively.
In (a)-(d), the black and blue circles are the experimental results of $\theta_0=\pi/6$ and $\pi/4$, respectively.
The associated solid lines are the results of an ideal evolution~\cite{SI}.}
\label{fig:fig2}
\end{figure}

To further illustrate the accumulation of the Berry phase,
we modify the external microwave signal and inspect the trajectory of the $\ket{s_\uparrow(t)}$ state
in the STA protocol. Without the excitation of the $\pi/2$-pulse, the ground-state qubit
(or the $\ket{s_\uparrow(t)}$ state)
is driven by the ramping-up field $\bm B_{\mathrm{ramp, tot}}(t_1)$ and
the subsequent $\mathcal C_+$-rotating field $\bm B_\mathrm{rot, tot}(t_2)$
with $T_{\mathrm{rot}}=30$ ns.
We measure the density matrix of the qubit by interrupting the rotation and
performing the QST every 0.5 ns. Figure~\ref{fig:fig2} presents the time
evolution of the qubit vector during the rotation period
for two pre-designed polar angles, $\theta_0=\pi/6$ and $\pi/4$, in the parameter space.
As shown in Figs.~\ref{fig:fig2}(c) and \ref{fig:fig2}(d), the polar angles of
both qubit vectors vary weakly with time (around the given value of $\theta_0$),
and their azimuthal angles increase almost linearly with the time as $\phi(t_2)=\omega_0 t_2$.
The time evolution of the radii in Fig.~\ref{fig:fig2}(b) shows
that the two qubit vectors are initially on the surface of the Bloch sphere
and slightly shrink due to dissipation.  Our experiment confirms that
the $\ket{s_\uparrow(t)}$ state follows the same direction of $\bm B_\mathrm{rot, 0}(t_2)$
and takes a circular path along the latitude
of $\pi/2-\theta_0$ in the northern hemisphere.
An integration, $\mathcal S=\int [1-\cos\theta(t_2)]d\phi(t_2)$,
further allows us to estimate the actual solid angles enclosed by the circular path.
The approximate integration results, $\mathcal S(\theta_0=\pi/6)=0.788$ and $\mathcal S(\theta_0=\pi/4)=1.72$,
are roughly close to their theoretical values of $\mathcal S=2\pi(1-\cos\theta_0)$
in an ideal adiabatic scenario, also consistent with the measurement of
the Berry phase in Fig.~\ref{fig:fig1}.

\begin{figure}
\centering
\includegraphics[width=0.65\columnwidth]{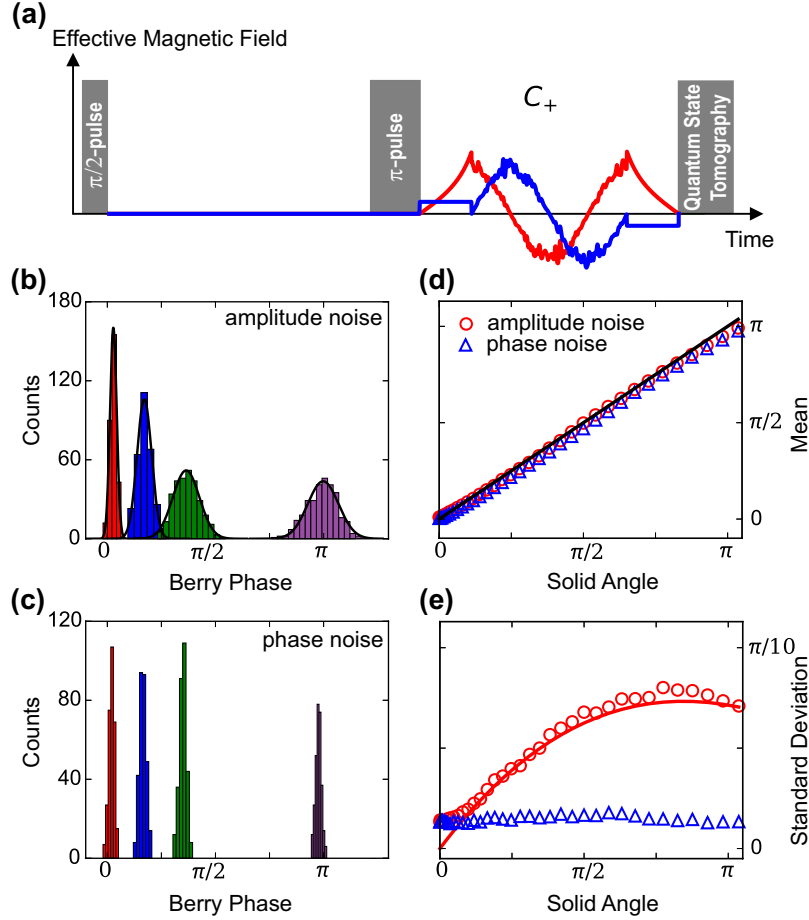}
\caption{The STA experiment of studying the noise influence on the measurement of the Berry phase.
(a) The schematic diagram of the effective magnetic field with a noisy $\mathcal C_+$-rotation.
The red and blue colors refer to the $x$- and $y$-components of the total control field.
With $T_\mathrm{rot}=30$ ns, the histograms of the Berry phase subject
to 300 trajectories of the amplitude ($c_\Omega=0.1$) and phase noises ($c_\phi=0.1$) are plotted in (b) and (c), respectively.
From the left to right in both (b) and (c), the four distributions refer to the results of
$\mathcal S=\pi/40, 3\pi/16, 3\pi/8$, and $\pi$ in the parameter space.
In (b), the solid lines are the Gaussian fitting curves of the histograms.
The mean value and standard deviation are shown in (d) and (e),
where the circles and up-triangles refer to the
experimental results of the amplitude and phase noises, respectively. In (d), the solid line is
the result without noise. In (e), the solid line is the analytical prediction of Eq.~(\ref{eq:5}).
}
\label{fig:fig3}
\end{figure}

In a previous experiment of a superconducting transmon qubit with the adiabatic protocol,  the stability of the Berry phase was studied
under the influence of slowly-varying noises~\cite{NoisePhase,NoiseTheory}. Similarly,
we introduce an artificial fluctuation $\delta \bm B_\mathrm{rot}(t)$
to the rotation field $\bm B_\mathrm{rot, tot}(t)$.
Two independent stochastic noises, $\delta \Omega(t)$ and $\delta \phi(t)$,
are assigned to the drive amplitude and phase of the effective magnetic field in the $x$-$y$ plane, respectively.
Both noises are assumed to follow the Ornstein-Uhlenbeck (O-U) process.
The amplitude noise satisfies $\langle \delta\Omega(t)\rangle =0$ and $\langle\delta\Omega(t)\delta\Omega(0)\rangle=c^2_\Omega \Omega^2_{\mathrm{tot}} \exp(-\Gamma t)$
with $\Omega_\mathrm{tot} = \Omega_0+\Omega_{\mathrm{cd}}$, while
the phase noise satisfies $\langle \delta\phi(t)\rangle =0$ and $\langle\delta\phi(t)\delta\phi(0)\rangle = c_\phi^2\exp(-\Gamma t)$.
The dimensionless coefficients, $c_\Omega$ and $c_\phi$, represent the reduced noise strengths.
A small noise bandwidth, $\Gamma=10$ MHz, is chosen
so that noises are correlated through the operation time.
In our experiment, we investigate the influence of the amplitude noise $\delta \Omega(t)$
and the phase noise $\delta \phi(t)$ separately. With respect to each noise parameter ($c_\Omega$ and $c_\phi$),
the total 300 stochastic trajectories of $\delta \bm B_\mathrm{rot}(t)$ are generated.
Since the relative dynamic phase between the $\ket{s_\uparrow(t)}$ and $\ket{s_\downarrow(t)}$ states
cannot be cancelled by a noisy spin-echo sequence, we only assign
a single rotating field, $\bm B_{\mathrm{rot, tot}}(t)+\delta \bm B_\mathrm{rot}(t)$,
to measure the Berry phase accumulated in a single cycle.
To reduce the influence of the intrinsic noise,
the spin-echo scheme is still used, while the effective magnetic field
is only applied in the second part following the $\pi$-pulse. An example
of the noisy magnetic field with the $\mathcal C_+$ rotation is shown in Fig.~\ref{fig:fig3}(a).
For each noisy sequence, we measure the final density matrix by the QST
and unwrap the total accumulated phase, similar to the method in Fig.~\ref{fig:fig1}.
The final relative Berry phase $\gamma$ ($|s_\downarrow(t)\rangle$ relative to $|s_\uparrow(t)\rangle$)
from one circular loop is then estimated from the total phase subtracted by
a theoretical calculation of the dynamic phase~\cite{SI}.
This indirect approach could introduce a small error of the dynamic phase
into the estimation of $\gamma$, which however does not affect the main conclusion
of our experiment. After collecting data over 300 trajectories, we
obtain the statistics of $\gamma$ subject to the amplitude noise $\delta \Omega(t)$
or the phase noise $\delta \phi(t)$.

With a fixed rotation time, $T_\mathrm{rot}=30$ ns,  we first study the variation of the Berry phase
under a given noise strength, $c_\Omega=0.1$ or $c_\phi=0.1$. For four pre-designed
solid angles, $\mathcal S=\pi/40, 3\pi/16, 3\pi/8$, and $\pi$ in the parameter space,
the histograms of $\gamma$  subject to the amplitude
and phase noises are presented in Figs.~\ref{fig:fig3}(b) and \ref{fig:fig3}(c), respectively.
For the amplitude noise, each histogram is well fitted by a Gaussian distribution, $\exp[-(\gamma-\bar{\gamma}_\Omega)^2/2\sigma^2_\Omega]$,
predicted by a perturbation theory~\cite{SI,NoiseTheory}.
As verified  in Fig.~\ref{fig:fig3}(d), the mean value of the Berry phase over various noise trajectories, $\bar{\gamma}_\Omega$,
is very close to the result of $\gamma=\mathcal S$
from a single circular loop without noise.  In the leading order of $c_\Omega$, the standard deviation $\sigma_{\Omega}$ is analytically expressed as~\cite{SI}
\begin{eqnarray}
\sigma_\Omega = 2\sqrt{2} c_\Omega \pi\sin^2\theta_0\cos\theta_0
\frac{\sqrt{\Gamma T_\mathrm{rot} - 1 + e^{-\Gamma T_\mathrm{rot}}}}{\Gamma T_\mathrm{rot}}.
\label{eq:5}
\end{eqnarray}
Figure~\ref{fig:fig3}(e) shows that the experimental measurement of $\sigma_\Omega$
agrees  well with Eq.~(\ref{eq:5}) over a broad range of $\mathcal S$.
At small solid angles, a residue $\sim0.02\pi$ is observed for $\sigma_\Omega$,
which may be attributed to the intrinsic noise of the qubit.
Equation~(\ref{eq:5}) indicates an upper deviation limit, $\sigma_\Omega\sim2 c_\Omega \pi\sin^2\theta_0\cos\theta_0$,
for the STA process. For the parameters $\Gamma T_\mathrm{rot}=0.3$ and $c_\Omega=0.1$ in our experiment,
Fig.~\ref{fig:fig3}(e) demonstrates that the fluctuation of the Berry phase is still tolerable with $\sigma_\Omega<0.1\pi$.
As a comparison, the influence of the phase noise is much weaker than that of the amplitude noise, which can be
identified by the narrow distributions of $\gamma$ in Fig.~\ref{fig:fig3}(c). This behavior is due to the fact that
the Berry phase depends on the geometry of the circular path instead of the rotating speed~\cite{NoiseTheory}. Accordingly,
the mean $\bar{\gamma}_\phi$ is close to the result without noise [Fig.~\ref{fig:fig3}(d)]
and the standard deviation $\sigma_\phi$ is consistently small, $\sigma_\phi\sim0.02\pi$, over the whole
range of the solid angle [Fig.~\ref{fig:fig3}(e)].

\begin{figure}
\centering
\includegraphics[width=0.65\columnwidth]{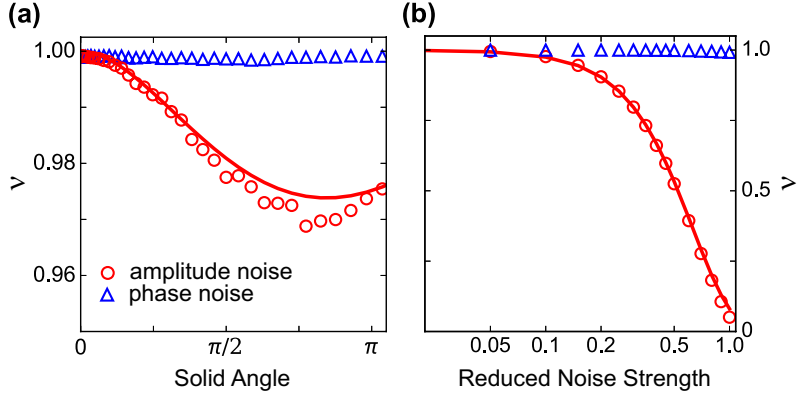}
\caption{ The coherence parameter $\nu$ under two types of noises with the rotation time $T_\mathrm{rot}=30$ ns.
(a) The results of $\nu$ versus the solid angle $\mathcal S$ for the noise strengths of $c_\Omega=0.1$ and $c_\phi=0.1$.
(b) The results of $\nu$ versus $c_\Omega$ and $c_\phi$ for $\mathcal S=\pi$.
In both (a) and (b), the circles and up-triangles refer to the experimental results of the amplitude
and phase noises, respectively. For the amplitude noise, the two solid lines are the results of
$\nu_\Omega=\exp(-\sigma^2_\Omega/2)$ with $\sigma_\Omega$ calculated by Eq.~(\ref{eq:5}).
}
\label{fig:fig4}
\end{figure}

A coherence parameter,  $\nu=|\left\langle \exp(i\gamma)\right\rangle|$, can alternatively
quantify the fluctuation of the Berry phase~\cite{NoisePhase}. For the amplitude noise
($c_\Omega=0.1$) and phase noise ($c_\phi=0.1$),
the $\mathcal S$-dependencies of $\nu_\Omega$ and $\nu_\phi$ in Fig.~\ref{fig:fig4}(a)
are consistent with the results of $\sigma_\Omega$ and $\sigma_\phi$ in Fig.~\ref{fig:fig3}(e).
The effect of the phase noise is essentially negligible whereas the influence of the amplitude noise is well described
by an analytical expression.
Furthermore, we explore the change of the coherence parameter as the noise strength ($c_\Omega$ or $c_\phi$) varies.
For a fixed polar angle, $\theta_0=\pi/3$ (or $\mathcal S=\pi$),
Fig.~\ref{fig:fig4}(b) confirms the weak influence of the phase noise.
As a comparison, $\nu_\Omega$ can maintain a large value for $c_\Omega<0.2$
while a major drop of $\nu_\Omega$ occurs for $c_\Omega>0.5$.
In both Figs.~\ref{fig:fig4}(a) and \ref{fig:fig4}(b), the results of $\nu_\Omega$ can be well
described by $\nu_\Omega=\exp(-\sigma^2_\Omega/2)$, where the analytical values of $\sigma_\Omega$ from Eq.~(\ref{eq:5}) are used.

In summary, we have implemented the STA protocol in a superconducting phase qubit.
In good agreement with the theoretical prediction,
the Berry phase is successfully measured in a time ($20~\mathrm{ns}\le T_\mathrm{rot}\le 60$ ns)
much shorter than that required by the adiabatic theorem ($T_\mathrm{rot}\gtrsim 1~\mu$s).
The measurement of the Berry phase is almost independent of the operation time, which is a characteristic
property of the STA protocol. The trajectory of a qubit state is verified, from which
the solid angle enclosed by the path is calculated to understand
the accumulation of the Berry phase.  Classical fluctuations of the drive amplitude or phase are
artificially introduced to the total control field. Our experiment shows that
the mean value of the Berry phase is unchanged,
while the standard deviation with the amplitude noise can be described by an analytical expression.
To further understand the stability of the Berry phase,
an extended study of quantum noise will be necessary in the future \cite{DMTong04,JFDu08}.
The fast measurement of the Berry phase in this paper will hopefully  encourage
more applications of the STA protocol in superconducting qubit systems.

\begin{acknowledgments}
This work is supported by the National Basic Research Program
of China (2014CB921203, 2015CB921004), the National Natural Science Foundation of
China (NSFC-11374260, NSFC-21173185), and the Fundamental Research Funds
for the Central Universities in China. Devices were made at John Martinis's group
using equipments of UC Santa Barbara Nanofabrication Facility, a part of the NSF-funded National
Nanotechnology Infrastructure Network.

\end{acknowledgments}

\end{document}


\title{Supplementary Material: Measuring the Berry Phase in a Superconducting Phase Qubit by a Shortcut to Adiabaticity}
\author{Zhenxing Zhang}
\author{Tenghui Wang}
\author{Liang Xiang}
\author{Jiadong Yao}
\author{Jianlan Wu}
\affiliation{Department of Physics, Zhejiang University, Hangzhou 310027, China}
\author{Yi Yin}
\email{yiyin@zju.edu.cn}
\affiliation{Department of Physics, Zhejiang University, Hangzhou 310027, China}
\affiliation{Collaborative Innovation Center of Advanced Microstructures, Nanjing 210093, China}

\maketitle

\section{Superconducting Phase Qubit Device and the Generation of Microwave Drive Signal}
\label{sec1}

\begin{figure}
\centering
\includegraphics[width=0.8\columnwidth]{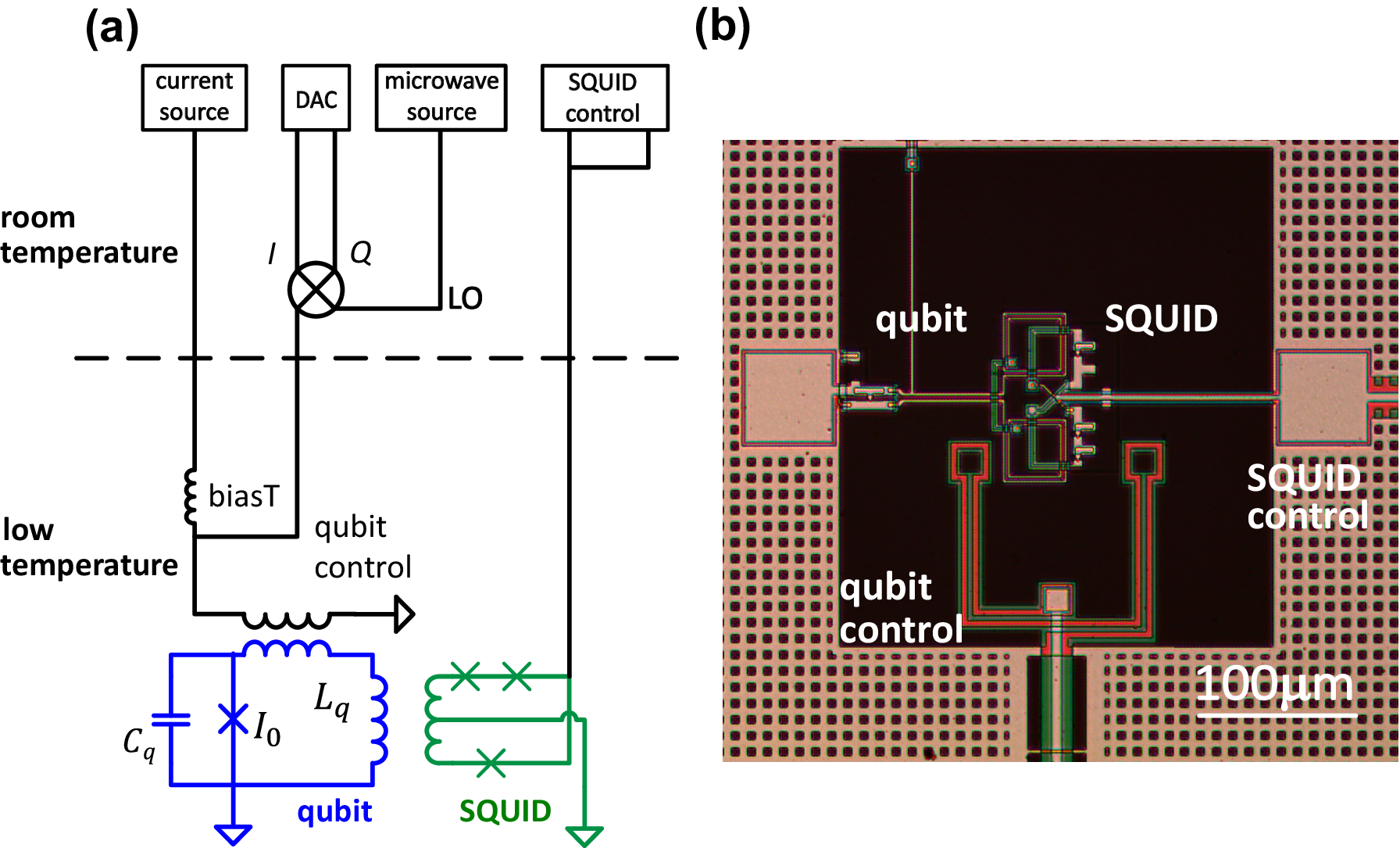}
\caption{(a) Schematic diagram of our experimental setup including the room-temperature control and the low-temperature phase qubit.
         (b) The optical micrograph of the phase qubit.
}
\label{fig:s1}
\end{figure}

Figure~\ref{fig:s1}(a) displays a schematic diagram of our experimental setup, including a phase qubit
and external control lines~\cite{MartinisReview}. The control signals are synthesized at room temperature, and then sent down to the low-temperature
stage (inside a dilution refrigerator whose base temperature is $\sim 10$ mK)
to manipulate and measure the qubit state. For the phase qubit placed in the dilution refrigerator, the main
components are a qubit, a superconducting quantum interference device (SQUID) and their
control lines. The optical micrograph of the phase qubit is shown in Fig.~\ref{fig:s1}(b).
The phase qubit is comprised of a Josephson junction (with a critical current $I_0$ = 2 $\mu$A), a parallel loop
inductance ($L_\mathrm{q}$ = 720 pH), and a capacitance ($C_\mathrm{q}$ = 1 pF).
The qubit control signal combines the flux current bias
and the microwave drive through a bias-tee.
The former signal from the current source sets the qubit resonance frequency,
while the latter signal drives the qubit state. A detailed description of the microwave drive signal
is provided in the next paragraph.
At the end of a quantum operation, the qubit state is projected to either the ground ($|0\rangle$) or
excited ($|1\rangle$) state for the readout measurement~\cite{MartinisReview}. As the ground and excited states induce  different fluxes in the qubit loop,
the SQUID can detect the probability of the two states through the SQUID control line.
In particular, the quantum state tomography (QST) technique is applied in the readout to
extract the density matrix of the final qubit state.

To describe the generation of a microwave drive signal $\lambda(t)$, we start with the
time-dependent Hamiltonian $H^{(\mathrm{S})}(t)$ in the Schrodinger picture (or the lab frame).
The general form of $H^{(\mathrm{S})}(t)$ is written as,
\be
H^{(\mathrm{S})}(t) &=& \hbar\omega_{10} |1\rangle\langle 1|+ \hbar \lambda(t) (|0\rangle\langle 1|+|1\rangle\langle 0|) \no \\
&=& \hbar\omega_{10} |1\rangle\langle 1|+ \hbar \Omega(t)\cos[\omega_\mathrm{d}t +\Phi(t)] (|0\rangle\langle 1|+|1\rangle\langle 0|),
\label{eq_S1}
\ee
where $\omega_{10}$ is the resonance  frequency of the qubit and $\omega_\mathrm{d}$  is the
drive frequency of a local oscillator (LO) [Fig.~\ref{fig:s1}(a)].
The LO signal is provided by a single microwave source
in our experiment.
The two high frequencies, $\omega_{10}$ and $\omega_\mathrm{d}$, are in the magnitude of GHz.
In addition, $\Omega(t)$ is  a drive amplitude in the unit of angular
frequency and $\Phi(t)$ is a time-varying phase.
An IQ mixer  mixes two low-frequency quadratures, $I(t)$ and $Q(t)$,
with the LO signal, producing an output signal,
$\lambda(t) = I(t)\cos\omega_\mathrm{d} t-Q(t)\sin\omega_\mathrm{d}t$.
To generate the microwave drive as in Eq.~(\ref{eq_S1}), the two quadratures are given by
\be
I(t) = \Omega(t)\cos\Phi(t),~~~\mathrm{and}~~~Q(t) = \Omega(t)\sin\Phi(t),
\label{eq_S3}
\ee
which are realized by two digital-to-analog-converter (DAC) outputs [Fig.~\ref{fig:s1}(a)].

\section{Rotating Frame of the External Field}
\label{sec2}

We introduce a rotating frame of the external field, in which the phase qubit can
be regarded as a spin-$1/2$ particle driven by an  effective magnetic field $\bm B(t)$.
In Eq.~(\ref{eq_S1}), the phase $\Phi(t)$ in the microwave drive signal $\lambda(t)$
is separated into two parts, $\Phi(t)=\xi(t)-\phi(t)$,
where $\phi(t)$ is considered as the phase in the $x$-$y$ plane.
The additional phase $\xi(t)$ is used to construct a time-varying drive frequency,
$\omega_\mathrm{d}+\delta\omega_\mathrm{d}t$ with $\delta\omega_\mathrm{d} (t)=\partial_t\xi(t)$.
Next we introduce a rotating-reference Hamiltonian,
$H_\mathrm{r}(t) = \hbar[\omega_\mathrm{d}+\delta\omega_\mathrm{d}(t)]|1\rangle\langle1|$,
and its time propagator, $U_\mathrm{r}(t)=\exp[-(i/\hbar) \int_0^t H_\mathrm{r}(\tau)d\tau]$.
The rotating frame with the reference frequency $\omega_\mathrm{d}+\delta\omega_\mathrm{d}(t)$ is built
in the interaction picture, where the Hamiltonian and the system
wavefunction are transformed into  $H^{(\mathrm{R})}(t) = U_\mathrm{r}^\dag(t) [H^{(\mathrm{S})}(t) -H_\mathrm{r}(t)] U_\mathrm{r}(t)$ and
 $|\psi^{(\mathrm{R})}(t)\rangle = U_\mathrm{r}^\dag(t)|\psi^{(\mathrm{S})}(t)\rangle$.
In particular, the Hamiltonian in the rotating frame is written as
\be
H^{(\mathrm{R})}(t)
&=&-\hbar \Delta(t) |1\rangle\langle 1|+ \hbar\lambda(t) e^{-i [\omega_\mathrm{d}t+\xi(t)]}|0\rangle\langle 1|+\hbar\lambda(t) e^{i [\omega_\mathrm{d}t +\xi(t)]}|1\rangle\langle 0|.
\label{eq_S15}
\ee
where $\Delta(t)=\omega_\mathrm{d}-\omega_{10}+\delta\omega_\mathrm{d}(t)$ is the detuning fluctuated around
a fixed number, $\Delta_0=\omega_\mathrm{d}-\omega_{10}$. By expressing the microwave drive signal as
\be
\lambda(t)=\frac{\Omega(t)}{2}\left\{\exp[i \omega_\mathrm{d}t +i\Phi(t)]+\exp[-i \omega_\mathrm{d}t -i\Phi(t)]\right\},
\ee
we take the rotating wave approximation (RWA) and ignore fast oscillations around $2\omega_\mathrm{d}$. As a result,
Eq.~(\ref{eq_S15}) is simplified to be
\be
H^{(\mathrm{R})}(t) &=&-\hbar \Delta(t) |1\rangle\langle 1|+ \frac{\hbar\Omega(t)}{2} e^{-i \phi(t)}|0\rangle\langle 1|+\frac{\hbar\Omega(t)}{2}  e^{i \phi(t)}|1\rangle\langle 0|.
\label{eq_S04}
\ee
The introduction of Pauli operators, $\sigma_x = |0\rangle\langle 1|+|1\rangle\langle 0| $,
$\sigma_y = -i|0\rangle\langle 1|+i|1\rangle\langle 0|$, and $\sigma_z = |0\rangle\langle0| - |1\rangle\langle1|$,
allows us to rewrite Eq.~(\ref{eq_S04}) as
\be
H^{(\mathrm{R})}(t) &=&\frac{\hbar}{2} \left[-\Delta(t) I + \Omega(t)\cos\phi(t) \sigma_x+ \Omega(t)\sin\phi(t) \sigma_y+ \Delta(t)\sigma_z\right],
\label{eq_S04a}
\ee
where $I=|0\rangle\langle 0|+|1\rangle\langle 1|$ is a unitary operator. After a energy shift of $-\hbar\Delta(t)/2$ for both the ground and excited states,
the Hamiltonian in the rotating frame is expressed in a vector form,
$H^{(\mathrm{R})}(t) = \hbar\bm B(t)\cdot \bm\sigma/2$, where
$\bm{\sigma}= (\sigma_x, \sigma_y,\sigma_z)$ is the vector of Pauli  operators, and
\be
\bm{B}(t) = (\Omega(t)\cos\phi(t), \Omega(t)\sin\phi(t),  \Delta(t))
\label{eq_S05}
\ee
is an effective magnetic field in the unit of angular frequency.

In our main text, we start the discussion from the adiabatic process in the rotating frame with a fixed reference frequency   $\omega_\mathrm{d}$, which implies a
fixed detuning $\Delta_0$ in the above derivation.
However, the counter-diabatic Hamiltonian in the `shortcut-to-adibabticity' (STA) protocol
induces a time-varying detuning $\Delta(t)$, which needs to be realized in the rotating frame with the reference frequency
$\omega_\mathrm{d}+\delta\omega_\mathrm{d}(t)$. For a Hamiltonian $H^{(\mathrm{R})}(t)$ which is the same in the two rotating frames,
the counterparts transformed in the lab frame
are however different, i.e.,
\be
H_1^{(\mathrm S)}(t) =  U_\mathrm{r}(\omega_\mathrm{d}; t) H^{(\mathrm{R})}(t) U^\dag_\mathrm{r}(\omega_\mathrm{d}; t) + H_\mathrm{r}(\omega_\mathrm{d}; t)
\label{eq_S06}
\ee
from the rotating frame of $\omega_\mathrm{d}$ and
\be
H_2^{(\mathrm S)}(t) =  U_\mathrm{r}(\omega_\mathrm{d}+\delta\omega_\mathrm{d}(t); t) H^{(\mathrm{R})}(t) U^\dag_\mathrm{r}(\omega_\mathrm{d}+\delta\omega_\mathrm{d}(t); t) + H_\mathrm{r}(\omega_\mathrm{d}+\delta\omega_\mathrm{d}(t); t)
\label{eq_S07}
\ee
from the rotating frame of $\omega_\mathrm{d}+\delta\omega_\mathrm{d}(t)$. In Eqs.~(\ref{eq_S06}) and (\ref{eq_S07}), the frequencies involved in
$H_\mathrm{r}(t)$ and $U_\mathrm{r}(t)$ are explicitly provided to clarify the difference of the two Hamiltonians.
Next we define the time propagators, $U^{(\mathrm S/\mathrm R)}(t)=T_+\exp[-(i/\hbar)\int_0^t H^{(\mathrm S/\mathrm R)}(\tau)d\tau]$,
where $T_+$ is the forward time ordering operator and the superscript $\mathrm S$ ($\mathrm R$) denotes the
lab (rotating) frame. For a given initial state $|\psi^{(\mathrm S)}(0)\rangle$ , the
system states in the lab frame are
\be
|\psi_1^{(\mathrm S)}(t)\rangle &=& U_\mathrm{r}(\omega_\mathrm{d}; t) U^{(\mathrm R)}(t) |\psi^{(\mathrm S)}(0)\rangle,
\label{eq_S08}
\ee
and
\be
 |\psi_2^{(\mathrm S)}(t)\rangle &=& U_\mathrm{r}(\omega_\mathrm{d}+\delta\omega_\mathrm{d}(t);t)U^{(\mathrm{R})}(t)|\psi^{(\mathrm{S})}(0)\rangle,
 \label{eq_S09}
\ee
with respect to the two Hamiltonians in Eqs.~(\ref{eq_S06}) and (\ref{eq_S07}), respectively.
In deriving Eqs.~(\ref{eq_S08}) and (\ref{eq_S09}), the relation,  $U^{(\mathrm S)}(t)= U_\mathrm{r}(t) U^{(\mathrm R)}(t)$,
is used, which then leads to
\be
|\psi_1^{(\mathrm S)}(t)\rangle &=&
U^\dag_\mathrm{r}(\delta\omega_\mathrm{d}(t);t)|\psi_2^{\mathrm{(S)}}(t)\rangle \no\\
&=& \big\{\langle 0|\psi_2^{(\mathrm S)}(t)\rangle\big\}|0\rangle + e^{i\xi(t)}\big\{\langle 1|\psi_2^{(\mathrm S)}(t)\rangle\big\} |1\rangle.
\label{eq_S11}
\ee
The phase shift $\exp[i\xi(t)]$ of the excited state is included in the QST,
so that our experiment based on $H_2^{(\mathrm S)}(t)$ in the lab frame can be used to study the STA protocol
in the rotating frame of $\omega_\mathrm{d}$.
This rotating frame will be used throughout the rest of the Supplementary Material
and the main text.
To simplify the notation, we will drop the superscript R for the rotating frame
and map the two-level system into a spin-$1/2$ particle by omitting the term $-\hbar\Delta(t) I/2$ in Eq.~(\ref{eq_S04a}).

\section{Derivation of The `Shortcut-to-Adiabaticity' Protocol}
\label{sec3}

Here we provide a theoretical derivation of the STA protocol, which is slightly different from the original one in Ref.~\cite{Berry09} but leads to the same result.
For a general quantum system, we consider a non-degenerate time-dependent Hamiltonian $H_0(t)=\sum_n \varepsilon_n(t)|n(t)\rangle\langle n(t)|$,
where $|n(t)\rangle$ is the $n$th instantaneous eigenstate associated with the eigenenergy $\varepsilon_n(t)$.
Each wavefunction can be linearly decomposed into $|\psi(t)\rangle = \sum_{n} a_n(t)|n(t)\rangle$ with $a_n(t)$ the time-dependent
coefficient. Following the Schrodinger equation, the time evolution of $a_n(t)$ is given by
\be
\hbar\dot{a}_n(t) = -i \left[ \varepsilon_n(t)-i\hbar\langle n(t)|\partial_tn(t)\rangle\right]a_n(t) - \hbar\sum_{m(\not=n)} \langle n(t)|\partial_t m(t)\rangle a_m(t).
\label{eq_S38}
\ee
In the adiabatic limit, the second term on the right hand side of Eq.~(\ref{eq_S38}) vanishes, resulting in
\be
\hbar\dot{a}_n(t) = -i \left[ \varepsilon_n(t)-i\hbar\langle n(t)|\partial_t n(t)\rangle\right]a_n(t).
\label{eq_S39}
\ee
The amplitude of $a_n(t)$ is unchanged with time and only a phase is accumulated, i.e., $a_n(t)=\exp[i \varphi_n(t)]a_n(0)$.

However, the influence from other eigenstates $|m(t)\rangle$ cannot be ignored if the time propagation of
$H_0(t)$ is not slow enough. To achieve a fast `adiabaticity', the STA protocol was proposed through the assistance of
a counter-diabatic Hamiltonian $H_{\mathrm{cd}}(t)$.
For the total Hamiltonian, $H_{\mathrm{tot}}(t)=H_0(t)+H_{\mathrm{cd}}(t)$, the wavefunction,
$|\psi(t)\rangle = \sum_{n} a_n(t)|n(t)\rangle$, is still decomposed using the eigen basis set of
the reference Hamiltonian $H_0(t)$. The time evolution of $a_n(t)$ is changed to be
\be
\hbar\dot{a}_n(t) &=& -i \left[ \varepsilon_n(t)-i\hbar\langle n(t)|\partial_t n(t)\rangle \right]a_n(t) - i  \langle n(t) |H_{\mathrm{cd}}(t)|n(t)\rangle a_n(t) \no \\
& & -i\sum_{m(\not=n)} \left[-i\hbar\langle n(t)|\partial_t m(t)\rangle +\langle n(t)|H_{\mathrm{cd}}(t)|m(t)\rangle\right] a_m(t).
\label{eq_S40}
\ee
To recover the adiabatic time evolution in Eq.~(\ref{eq_S39}), the counter-diabatic Hamiltonian is required to satisfy
\be
\left\{\ba{ll}\langle n(t) |H_{\mathrm{cd}}(t)|n(t)\rangle = 0  &~~  \\
\langle n(t) |H_{\mathrm{cd}}(t)|m(t)\rangle = i\hbar\langle n(t)|\partial_t m(t)\rangle &~~~\mathrm{for}~~~m\neq n \ea  \right. .
\label{eq_S41}
\ee
Since the indices, $m$ and $n$, are arbitrary, the action of $H_\mathrm{cd}(t)$ applied to each $|n(t)\rangle$
must follow $H_{\mathrm{cd}}(t)|n(t)\rangle =i \hbar[|\partial_t n(t)\rangle- \langle n(t)|\partial_t n(t)\rangle |n(t)\rangle]$.
The counter-diabatic Hamiltonian is thus given by
\be
H_{\mathrm{cd}}(t) = i \hbar\sum_n \big[ |\partial_t n(t)\rangle- \langle n(t)|\partial_t n(t)\rangle |n(t)\rangle \big] \langle n(t)|,
\label{eq_S42}
\ee
which satisfies $\sum_{m, n}H^\ast_{0; m, n}(t)H_{\mathrm{cd}; m, n}(t)=0$.

\section{The Berry Phase of a Two-Level System with the STA Protocol}
\label{sec4}

Here we derive the Berry phase
of a two-level system with the STA protocol. As demonstrated in Supplementary Material II, 
the two-level system can be mapped to a spin-$1/2$ particle. The reference Hamiltonian is
represented in a general form, $H_0(t) = \hbar\bm{B}_0(t)\cdot{\bm\sigma}/2$,
where $\bm B_0(t)=(\Omega(t)\cos\phi(t), \Omega(t)\sin\phi(t),  \Delta(t))$ is an effective magnetic field in the rotating frame.
For simplicity, both $\Omega(t)$ (the amplitude in the $x$-$y$ plane) and $\Delta(t)$ (the detuning along the $z$-axis)
are assumed to be positive. The vector amplitude of the control field is given by
$B_0(t)=|\bm B_0(t)|=\sqrt{\Omega^2(t)+\Delta^2(t)}$.
In a normalized parameter sphere of $\bm B_0(t)/B_0(t)$, we introduce the
polar angle, $\theta(t)=\arctan[\Omega(t)/\Delta(t)]$, and the azimuthal angle (phase in the $x$-$y$ plane)
$\phi(t)$  to define  the spherical surface.

For this reference Hamiltonian $H_0(t)$,  its instantaneous
spin-up ($|s_\uparrow(t)\rangle$) and spin-down ($|s_\downarrow(t)\rangle$) states are expanded over the qubit
states ($|0\rangle, |1\rangle$),
\be
\left\{\ba{lll}
|s_\uparrow(t)\rangle &=& \cos\frac{\theta(t)}{2}|0\rangle + e^{i\phi(t)} \sin\frac{\theta(t)}{2}|1\rangle,   \\
|s_\downarrow(t)\rangle &=& - e^{-i\phi(t)}\sin\frac{\theta(t)}{2}|0\rangle+\cos\frac{\theta(t)}{2}|1\rangle.
\ea \right. \label{eq_S44}
\ee
The reference Hamiltonian is recast into $H_0(t) = \sum_{n=\uparrow, \downarrow}\varepsilon_n(t) |s_n(t)\rangle\langle s_n(t)|$,
with the instantaneous eigenvalues $\varepsilon_{\uparrow, \downarrow}(t) = \pm\hbar B_0(t)/2$ .
The wavefunction is decomposed into $|\psi(t)\rangle = \sum_{n=\uparrow, \downarrow}a_n(t)|s_n(t)\rangle$, and 
Eq.~(\ref{eq_S44}) is rewritten as $|s_{n=\uparrow, \downarrow}(t)\rangle  = \sum_{i=\uparrow,\downarrow} b_{n, i}(t) |i\rangle$ in a simplified notation.
The counter-diabatic Hamiltonian in Eq.~(\ref{eq_S42}) is then written explicitly as
\be
H_{\mathrm{cd}}(t) = i\hbar\sum_{i, i^\pr=\uparrow,\downarrow}\left[\sum_n \partial_t b_{n,i}(t) b^\ast_{n,i^\pr}(t)
-\sum_{n, j} \partial_t b_{n,j}(t) b^\ast_{n,j}(t) b_{n, i}(t)b^\ast_{n,i^\pr}(t)\right] |i\rangle\langle i^\pr|.
\label{eq_S46}
\ee
With the help of Pauli operators, Eq.~(\ref{eq_S46}) is organized into a simple form,
$H_{\mathrm{cd}}(t)= \hbar \bm B_\mathrm{cd}(t)\cdot \bm \sigma/2$, where the counter-diabatic effective magnetic
field is given by
\be
\left\{\ba{lll} B_{\mathrm{cd}; x}(t) &= & -\dot{\theta}(t)\sin\phi(t)-\dot{\phi}(t)\sin\theta(t)\cos\theta(t)\cos\phi(t) \\
                B_{\mathrm{cd}; y}(t) &= & \dot{\theta}(t)\cos\phi(t)-\dot{\phi}(t)\sin\theta(t)\cos\theta(t)\sin\phi(t) \\
                B_{\mathrm{cd}; z}(t) &= &\dot{\phi}(t)\sin^2\theta(t) \ea \right. .
\label{eq_S43}
\ee
In a vector representation, the counter-diabatic magnetic  field is equal to a cross product,
\be
\bm B_\mathrm{cd}(t) = \frac{1}{|\bm B_0(t)|^2}\bm B_0(t)\times\dot{\bm B}_0(t).
\label{eq_S48}
\ee

In the STA protocol, the time evolution of the two-level system becomes adiabatic with respect to the reference Hamiltonian.
The coefficients $a_{n=\uparrow, \downarrow}(t)$ of the two instantaneous eigenstates are governed by
\be
\dot{a}_n(t) &=& -i \left[\varepsilon_n(t)/\hbar-i \langle s_n(t)|\partial_t s_n(t)\rangle \right] a_n(t).  \label{eq_S49}
\ee
For each coefficient, a phase $\varphi_n(t)$ is accumulated with time and can be separated into two parts, $\varphi_n(t)=\alpha_n(t)+\gamma_n(t)$.
The first part, $\alpha_{n}(t) = -\frac{1}{\hbar}\int_0^t \varepsilon_{n}(\tau) d\tau$, relies on the time-dependent vector amplitude
$B_0(t)$ and is termed the dynamic phase. The second part,
$\gamma_n(t)= i \int_0^t \langle s_n(\tau)|\partial_\tau s_n(\tau)\rangle d\tau$, is a function of the
polar angle $\theta(t)$ and the azimuthal angle $\phi(t)$. A curve $\bm R(t)$ is defined
on the surface of the Bloch sphere (or the normalized parameter sphere)
by $\bm R(t)=\{\theta=\theta(t), \phi=\phi(t)\}$. The time differential in $\gamma_n(t)$ can be changed to the spatial gradient, giving
 $\gamma_n(t)= i \int_{\bm R(0)}^{\bm R(t)}  \langle n(\bm{R})|\nabla_{\bm{R}} n(\bm{R})\rangle \cdot d{\bm R}$. If the path $\bm R(t)$
 is closed after the time evolution, there is no explicit time dependence in the phase $\gamma_n(t)$, giving
\be
\gamma_n = i\oint_\mathcal C \langle n(\bm{R})|\nabla_{\bm{R}} n(\bm{R})\rangle \cdot d{\bm R}.
\label{eq_S51}
\ee
The phase $\gamma_n$ is considered as the Berry phase with respect to the reference Hamiltonian, even though
the fast STA protocol is applied. For the two-level system, the Berry phase in Eq.~(\ref{eq_S51})
can be further simplified to
\be
\gamma_{\uparrow, \downarrow} = \mp \frac{1}{2}\oint_\mathcal C [1-\cos\theta] d\phi.
\label{eq_S53}
\ee
where the signs $\pm$ refer to the instantaneous spin-up and spin-down states, respectively.
In the above definition of instantaneous eigenstates, we may
consider a gauge transformation, i.e., $|s_n(t)\rangle \rightarrow \exp[i \zeta_n(t)]|s_n(t)\rangle$.
After a straightforward re-derivation, we can demonstrate that a phase shift of $2k\pi$ ($k\in \mathrm{integers}$)
is allowed in the Berry phase,  i.e., $\gamma_n \rightarrow \gamma_n +2k\pi$. For convenience, the Berry phase
in our experiment is assumed to follow the result in Eq.~(\ref{eq_S53}) without an additional phase shift of $2k\pi$.

It is impossible to experimentally extract the absolute phase of a single quantum state.
One way of indirectly extracting the Berry phase
is to numerically calculate the solid angle,
$\mathcal S = \oint_\mathcal C [1-\cos\theta] d\phi$, by measuring the trajectory of the qubit vector on the Bloch sphere.
In a superconducting Cooper pair pump, the phase accumulation speed of the ground state
can be measured through the pumped charge, which also allows an estimation of the Berry phase~\cite{MottononPhase}.
Another approach lies on the measurement of the phase difference by preparing a
superposition of two instantaneous eigenstates.
In our experiment, a spin-echo scheme with the initial state $(|0\rangle+|1\rangle)/\sqrt{2}$
is applied. As the dynamic phase is removed by the spin-echo sequence,
the phase difference of $|1\rangle$ relative to $|0\rangle$ gives rise to the difference of the
Berry phase.

\section{The Berry Phase Subject to a Rotating Field}
\label{sec5}

In this Supplementary Material, we provide the theoretical prediction of
the Berry phase for the instantaneous spin-up state subject to a rotating field.

At the very begining of our experiment, the $|0\rangle$ and $|1\rangle$ states of the qubit are
the instantaneous spin-up ($|s_\uparrow(t)\rangle$) and spin-down ($|s_\downarrow(t)\rangle$) states
in the rotating frame, respectively.
Here we discuss the behavior of  $|s_\uparrow(t)\rangle$, and the opposite way is applied to $|s_\downarrow(t)\rangle$.
Since the Berry phase is not accumulated in the ramping-up and ramping-down steps
due to the fact that $\phi(t)$ is not changed, we focus on the two rotating steps,
where the reference magnetic field follows
\be
\bm B_{0}(t)= B_0(\sin\theta_0\cos\phi(t), \sin\theta_0\sin\phi(t), \cos\theta_0),
\ee
with $B_0=\sqrt{\Omega_0^2+\Delta_0^2}$ and $\theta_0 =\arctan(\Omega_0/\Delta_0)$.
As the system evolves in the instantaneous spin-up state,
the wavefunction is written as  $|\psi(t)\rangle = a_\uparrow(t) |s_\uparrow(t)\rangle$, where the
accumulated phase is included in the coefficient $a_\uparrow(t)$.
The system wavefunction $|\psi(t)\rangle$ is represented by a Bloch vector, which points to the same direction as the reference magnetic field $\bm B_0(t)$.
The trajectory of $|\psi(t)\rangle$ is characterized by
\be
r_\uparrow(t)=1,~~~\theta_\uparrow(t)=\theta_0, ~~~\mathrm{and}~~~ \phi_\uparrow(t)=\phi(t),
\label{eq_S61}
\ee
where $r_\uparrow(t)$, $\theta_\uparrow(t)$ and $\phi_\uparrow(t)$ are the radius,
polar and azimuthal angles on the Bloch sphere.
In our experiment, we consider a constant rotating speed $\omega_0$ along the counterclockwise ($\mathcal C_+$)
or clockwise ($\mathcal C_-$) direction, i.e., $\phi(t)=\pm \omega_0 t$.
If $|\psi(t)\rangle$ evolves over a single circular rotation,  we apply Eq.~(\ref{eq_S53})
to calculate the Berry phase,
\be
\gamma_\uparrow = \mp \pi (1-\cos\theta_0),
\label{eq_S58}
\ee
where the $\mp$ signs correspond to the counterclockwise and clockwise rotations, respectively.
Following the same approach, we can obtain the expressions of the spin-down state.

In the first part of our spin-echo scheme, the accumulated phases for the $|s_\uparrow(t)\rangle$
and $|s_\downarrow(t)\rangle$ are opposite, i.e., $\alpha_\downarrow(t)=-\alpha_\uparrow(t)$
and $\gamma_\downarrow = -\gamma_\uparrow$.
These two coefficients, $a_\uparrow(t)$ and $a_\downarrow(t)$, are swapped by a refocusing $\pi$-pulse.
The wavefunction is changed to
\be
|\psi(t)\rangle &\propto& e^{i\alpha_\downarrow(t)} e^{i\gamma_\downarrow}|s_\uparrow(t)\rangle+e^{i\alpha_\uparrow(t)} e^{i\gamma_\uparrow} |s_\downarrow(t)\rangle \no \\
&=& e^{-i\alpha_\uparrow(t)} e^{-i\gamma_\uparrow}|s_\uparrow(t)\rangle+ e^{-i\alpha_\downarrow(t)} e^{-i\gamma_\downarrow}|s_\downarrow(t)\rangle.
\ee
For each intantaneous eigenstate subject to the second part of the spin-echo sequence,
the dynamic phase is the same as that accumulated in the first part while the Berry phase is opposite
due to a reversed rotating direction.  At the echo time when the two instantaneous eigenstates return
to their initial positions ($|s_\uparrow(t)\rangle=|0\rangle$ and $|s_\downarrow(t)\rangle=|1\rangle$),
the wavefunction is given by
\be
|\psi(t)\rangle \propto e^{-2i\gamma_\uparrow} |0\rangle + e^{-2i\gamma_\downarrow}|1\rangle,
\ee
where $\gamma_{n=\uparrow, \downarrow}$ is the Berry phase from the one cycle in the first part.
The density matrix of this final qubit state, $\rho=|\psi(t)\rangle\langle\psi(t)|$,
is extracted by the QST. The phase difference, $\exp(i\gamma)=\langle 1|\rho| 0\rangle$,
is used to measure the Berry phase,
\be
\gamma =\mp 4\pi (1-\cos\theta_0),
\label{eq_S59}
\ee
where the $\mp$ signs refer to the $\mathcal C_{+-}$ and $\mathcal C_{-+}$ spin-echo procedures, respectively.

\section{Analytical Prediction for a Slowly-Varying Noise in the STA Process}
\label{sec7}

We apply a theoretical method, similar to the approach in Ref.~\cite{NoiseTheory},
to obtain an analytical expression for a slowly-varying classical noise
in the STA process.
For simplicity, we ignore the intrinsic relaxation and decoherence.
A classical Gaussian noise $\delta H(t)$ is considered for the total Hamiltonian, $H_\mathrm{tot}(t)=H_0(t)+H_\mathrm{cd}(t)$,
during the rotation period. The total rotating field without noise is given by
$\bm B_{\mathrm{tot}}(t)= (\Omega_\mathrm{tot}\cos\phi(t), \Omega_\mathrm{tot}\sin\phi(t), \Delta_\mathrm{tot})$,
where $\Omega_\mathrm{tot}$ and $\Delta_\mathrm{tot}$ include the modifications of the counter-diabatic field.
For convenience, we drop the symbol `rot' in the representation of an effective magnetic field in this Supplementary Material.
We consider three possible fluctuations, $\delta \Delta(t)$ for the detuning along the $z$ direction,
$\delta \Omega(t)$ and $\delta \phi(t)$ for the drive amplitude and phase in the $x-y$ plane.
Accordingly, the three types of stochastic rotating fields are explicitly written as
\be
\left\{\ba{lll}
\bm B_\mathrm{tot}(t)+\delta\bm B_{\Delta}(t)&=&(\Omega_\mathrm{tot}\cos\phi(t), \Omega_\mathrm{tot}\sin\phi(t), \Delta_\mathrm{tot}+\delta\Delta(t))  \\
\bm B_\mathrm{tot}(t)+\delta\bm B_\Omega(t)&=&([\Omega_\mathrm{tot}+\delta\Omega(t)]\cos\phi(t), [\Omega_\mathrm{tot}+\delta\Omega(t)]\sin\phi(t), \Delta_\mathrm{tot}) \\
\bm B_\mathrm{tot}(t)+\delta\bm B_\phi(t)  &=& (\Omega_\mathrm{tot}\cos[\phi(t)+\delta\phi(t)], \Omega_\mathrm{tot}\sin[\phi(t)+\delta\phi(t)], \Delta_\mathrm{tot})
\ea \right. .
\label{eq_S64}
\ee
Since the influences of $\delta\Delta(t)$ and $\delta \Omega(t)$ are similar,
we only apply  $\delta\bm B_\Omega(t)$ and $\delta\bm B_\phi(t)$  in our experiment.
The Ornstein-Uhlenbeck  process is assigned for $\delta \Omega(t)$ and $\delta \phi(t)$, giving
\be
\langle \delta\Omega(t)\rangle =0,~~~ \langle\delta\Omega(t)\delta\Omega(0)\rangle=c^2_\Omega \Omega^2_{\mathrm{tot}} \exp(-\Gamma t),
\ee
and
\be
\langle \delta\phi(t)\rangle =0, ~~~\langle\delta\phi(t)\delta\phi(0)\rangle = c_\phi^2\exp(-\Gamma t).
\ee
Here $c_\Omega$ and $c_\phi$ are the reduced noise strengths and $\Gamma$ is the noise bandwidth.

Next we discuss the behaviors of the two noises separately. ({\bf i}) The Berry phase accumulated in the rotating period
is not affected by the phase noise $\delta\phi(t)$. Since the polar angle is fixed at $\theta_0$ in the rotating step,
the Berry phase is simplified to $\gamma_{n=\uparrow,\downarrow} = \mp(1/2)(1-\cos\theta_0)\oint_\mathcal C d[\phi+\delta\phi]$.
If the  effective magnetic field $\delta\bm B_\Omega(t)$ undergoes a closed path,
the integration of $\delta\phi$ vanishes and the result of $\gamma_n$ is the same as that without noise.

({\bf{ii}}) For the influence of the amplitude noise,  we first assume that $\delta\Omega(t)$ slowly varies
with time (behaves similarly as a static disorder which is relevant in an adiabatic process). The fluctuated magnetic field,
$\bm B_\mathrm{tot}(t)+\delta\bm B_\Omega(t)$, can be factorized into a fluctuated reference field,
$\bm B_0(t)+\delta\bm B_{0; \Omega}(t)$, and its counter-diabatic correction. The first order expansion of $\delta\Omega(t)$,
gives rise to
\be
\bm B_0(t)+\delta\bm B_{0; \Omega}(t)&=& ([B_0(t)+\delta B_0(t)]\sin[\theta_0+\delta\theta(t)]\cos\phi(t), \no \\
&&~[B_0(t)+\delta B_0(t)]\sin[\theta_0+\delta\theta(t)]\sin\phi(t), \no \\
&&~[B_0(t)+\delta B_0(t)]\cos[\theta_0+\delta\theta(t)])
\ee
with
\be
\delta B_0(t)      &=& \sin\theta_0(\Omega_\mathrm{tot}  \mp  \omega_0\sin\theta_0\cos\theta_0)\frac{\delta\Omega(t)}{\Omega_\mathrm{tot}}+O(\delta\Omega^2(t)),  \label{eq_S65}\\
\delta \theta(t) &=& \sin\theta_0\cos\theta_0 \frac{ \delta\Omega(t)}{\Omega_\mathrm{tot}}+O(\delta\Omega^2(t)). \label{eq_S66}
\ee
 where the signs $\mp$ refer to the counterclockwise ($\mathcal{C}_+$) and clockwise ($\mathcal{C}_-$) directions, respectively.
For the example of a single $\mathcal{C}_+$-rotation, the dynamic  and Berry
phase differences of $|s_\downarrow(t)\rangle$ relative to $|s_\uparrow(t)\rangle$ are fluctuated, following
\be
\delta \alpha   &=& \sin\theta_0(\Omega_\mathrm{tot} - \omega_0\sin\theta_0\cos\theta_0) \int_0^{T_\mathrm{rot}} \frac{\delta\Omega(\tau)}{\Omega_\mathrm{tot}} d\tau +O(\delta\Omega^2(t)), \label{eq_S67}\\
\delta \gamma &= &  \omega_0\sin^2\theta_0\cos\theta_0 \int_0^{T_\mathrm{rot}}  \frac{\delta\Omega(\tau)}{\Omega_\mathrm{tot}} d\tau +O(\delta\Omega^2(t)),
\label{eq_S68}
\ee
respectively. In our experiment with a noisy pulse, we measure the total relative phase 
from the QST. It is however hard to directly extract $\gamma$ since the noise can destroy the cancellation of the dynamic phase in
the spin-echo scheme. An indirect approach is to record
the input noise $\delta\Omega(t)$ and theoretically calculate the relative dynamic phase, $\alpha+\delta\alpha$, for each noisy trajectory.
The corresponding relative Berry phase is estimated by $\gamma[\delta\Omega(t)]=\varphi[\delta\Omega(t)]-\alpha-\delta\alpha[\delta\Omega(t)]$,
where $\varphi[\delta\Omega(t)]$ is the total relative phase.
Based on the perturbed result in Eq.~(\ref{eq_S68}), the statistics of the fluctuated Berry phase is characterized by
the mean $\langle\delta \gamma\rangle=0$ and the standard deviation
\be
\sigma_\Omega^2 &=& \langle \delta \gamma^2 \rangle =  8 c^2_\Omega \pi^2\sin^4\theta_0\cos^2\theta_0  \frac{\Gamma {T_{\mathrm{rot}}} - 1 + \exp(-\Gamma {T_{\mathrm{rot}}})}{\Gamma^2T_{\mathrm{rot}}^2}.
\label{eq_S70}
\ee
A Gaussian distribution is expected for $\delta \gamma$ since the underlying noise $\delta\Omega(t)$ is Gaussian.
The alternative coherence parameter,  $\nu=|\langle\exp(i\gamma)\rangle|$, is fully determined by
the first and second moments of $\delta\gamma$, giving
\be
\nu &=& \exp(-\sigma^2_\gamma/2) \no \\
&=&\exp\left[-4 c^2_\Omega \pi^2\sin^4\theta_0\cos^2\theta_0  \frac{\Gamma {T_{\mathrm{rot}}} - 1 + \exp(-\Gamma {T_{\mathrm{rot}}})}{\Gamma^2T_{\mathrm{rot}}^2}\right].
\label{eq_S71}
\ee

\section{Estimation of a Geometric Phase Gate Fidelity with the STA Protocol}
\label{sec8}

The accumulated Berry phase can be utilized in the realization of a geometric phase gate.
In this Supplementary Material, we provide a numerical estimation on
the fidelity of a $\pi$-phase gate.

A unitary operation is performed onto the initial state $|\psi(0)\rangle$ in an
ideal quantum operation. The quantum state $|\psi(t_{\rm f})\rangle$
at the final time $t_{\rm f}$ is given by $|\psi(t_{\rm f})\rangle = U |\psi(0)\rangle$.
For the $\mathcal C_{+-}$ spin-echo procedure in our experiment,
the unitary operator $U$ is explicitly written as
\be
U = \left(\ba{cc} 0 & \exp[i\cal S] \\ \exp[-i\cal S] & 0 \ea \right),
\label{eq_S73}
\ee
where the global phase is excluded and ${\cal S} = 2\pi(1-\cos\theta_0)$ is the designed solid angle.
A subsequent $\pi_x$-pulse leads to the overall unitary operation,
\be
U_{\rm tot}=(-i\sigma_x) U=\exp[-i({\cal S}+\frac{\pi}{2})]\left(\ba{cc} 1 & 0 \\  0 & \exp[i2\cal S]\ea \right),
\label{eq_S73a}
\ee
which corresponds to a $2\cal S$-phase gate. In the special case of $\theta_0=\arccos(3/4)$,
we obtain a $\pi$-phase gate, i.e., $U_{\rm tot}\propto \sigma_z$.

\begin{table}[t!]
\begin{tabular}{|c|c|c|c|c|c|}\hline
  &  Protocol & $\Delta_0/2\pi$ &  $T_\mathrm{ramp}$ & $T_{\mathrm{rot}}$ & Fidelity\\ \hline
 phase qubit &    Adiabatic & 7 MHz & 350 ns & 1000 ns & 0.2500 \\ \cline{2-6}
( $T_1$ = 270 ns, $T_2^{\mathrm{echo}}$ = 450 ns)&  STA &  7 MHz  & 10 ns & 30 ns & 0.7023 \\ \hline
Xmon qubit &   Adiabatic & 7 MHz &350 ns & 1000 ns & 0.8465\\ \cline{2-6}
( $T_1$ = 20 $\mu$s, $T_2^{\mathrm{echo}}$ = 20 $\mu$s) &    STA & 7 MHz & 10 ns & 30 ns & 0.9936\\ \hline
\end{tabular}
\caption{The fidelities of the $\pi$-phase gate in our phase qubit and a typical Xmon qubit.
Both the STA and adiabatic protocols are studied. All the results are numerically obtained by the Lindbald simulation.
}
\label{tab_S1}
\end{table}

A practical quantum operation is limited by quantum dissipation. Here we use
the Lindblad equation,
\be
\partial_t \rho(t) &=& -\frac{i}{\hbar} [H(t), \rho(t)]+\frac{1}{T_1}\left[\sigma_- \rho(t)\sigma_+-\frac{1}{2}\sigma_+\sigma_-\rho(t)-\frac{1}{2}\rho(t)\sigma_+\sigma_- \right] \no \\
&&+\frac{2}{T^{\mathrm{echo}}_2}\left[\sigma_+\sigma_- \rho(t)\sigma_+\sigma_--\frac{1}{2}\sigma_+\sigma_-\sigma_+\sigma_-\rho(t)-\frac{1}{2}\rho(t)\sigma_+\sigma_-\sigma_+\sigma_- \right],
\label{eq_S74}
\ee
to numerically estimate the time evolution of the density matrix $\rho(t)$, where $\sigma_+=|1\rangle\langle0|$ and $\sigma_-=|0\rangle\langle1|$
are two Lindblad operators. A quantum dynamical map  is then defined between the initial and final density matrices ($\rho(0)$ and $\rho(t_{\rm f})$ respectively),
i.e.,
\be
\rho(t_{\rm f}) = \sum_{i, j=1}^4 \chi_{i, j} u_i \rho(0) u^\dag_j,
\label{eq_S75}
\ee
where the four operators of the SU(2) group, $\{u_1 =I, u_2 = \sigma_x, u_3 = \sigma_y, u_4 = \sigma_z\}$,
are used as the expansion bases. The $4\times 4$ $\chi$-matrix defined in Eq.~(\ref{eq_S75}) is 
independent of the initial density matrix $\rho(0)$. For an ideal $\pi$-phase gate, the $\chi$-matrix
satisfies $\chi^{\mathrm{ideal}}_{i, j} = \delta_{i, 4}\delta_{j, 4}$. The accuracy of a practical $\pi$-phase gate can be characterized by its gate fidelity,
given by~\cite{nonAbelianGate}
\be
F =  \mathrm{Tr} \left\{ \chi^{\mathrm{ideal}} \chi \right\}.
\label{eq_S76}
\ee
In Table~\ref{tab_S1}, we provide the numerical estimations of $F$ for our phase qubit ($T_1=270$ ns and $T^{\mathrm{echo}}_2=450$ ns)
and a typical Xmon qubit ($T_1=20~\mu$s and $T^{\mathrm{echo}}_2=20~\mu$s).
Both the STA ($T_{\mathrm{ramp}}=10$ ns and $T_{\mathrm{rot}}=30$ ns) and adiabatic ($T_{\mathrm{ramp}}=350$ ns and $T_{\mathrm{rot}}=1000$ ns) protocols
are considered. Our numerical results show that the STA protocol can help establish a higher fidelity in a  operation time
much shorter than that required by the adiabatic theroem.
It will be interesting to explore the STA protocol in the Xmon qubit (e.g., the STA $\pi$-phase gate with fidelity $>99\%$) in the future.